\documentclass[fleqn,usenatbib]{mnras}

\usepackage{newtxtext,newtxmath}

\usepackage[T1]{fontenc}
\usepackage{ae,aecompl}
\DeclareRobustCommand{\VAN}[3]{#2}
\let\VANthebibliography\thebibliography
\def\thebibliography{\DeclareRobustCommand{\VAN}[3]{##3}\VANthebibliography}

\usepackage{graphicx}	
\usepackage{amsmath}	
\usepackage{amssymb}	
\usepackage{geometry}
\usepackage{ulem}
\usepackage{tabularx,booktabs}
\usepackage{hyperref}
\usepackage{cleveref}
\usepackage{natbib}

\newcommand{\pcm}{\,cm$^{-2}$}	
\newcommand{\ps}{\,s$^{-1}$}	
\newcommand{\pcms}{\,cm$^{-2}$\,s$^{-1}$}	

\title[Spectroscopic studies of LMC X-1 using \textit{AstroSat}]{Estimation of the black hole spin in LMC X-1 using \textit{AstroSat}}
\author[S. P. Mudambi et al.]
{Sneha Prakash Mudambi,$^{1}$
A. Rao,$^{2}$
S. B. Gudennavar,$^{1}$\thanks{E-mail: shivappa.b.gudennavar@christuniversity.in (SBG); sneha.m@res.christuniversity.in (SPM); anjalirao115@gmail.com (AR)}
R. Misra,$^{3}$
and S. G. Bubbly,$^{1}$
\\\\
$^{1}$Department of Physics and Electronics, CHRIST (Deemed to be University), Bangalore Central Campus, Bengaluru-560029, India\\
$^{2}$Department of Physics and Astronomy, Faculty of Physical Sciences and Engineering, University of Southampton, Southampton S017 1BJ, UK\\ 
$^{3}$Inter-University Centre for Astronomy and Astrophysics, Ganeshkind, Pune-411007, India
}
\date{Accepted XXX. Received YYY; in original form ZZZ}
\pubyear{2020}
\begin{document}
\label{firstpage}
\pagerange{\pageref{firstpage}--\pageref{lastpage}}
\maketitle
\begin{abstract}
LMC X-1, a persistent, rapidly rotating, extra-galactic, black hole X-ray binary (BHXB) discovered in 1969, has always been observed in its high soft state. Unlike many other BHXBs, the black hole mass, source distance and binary orbital inclination are well established. In this work, we report the results of simultaneous broadband  spectral studies of LMC X-1 carried out using the data from Soft X-ray Telescope and Large Area X-ray Proportional Counter aboard \textit{AstroSat} as observed on 2016 November 26$^{th}$ and 2017 August 28$^{th}$. The combined spectrum was modelled with a multicolour blackbody emission (\textit{diskbb}), a \textit{Gaussian} along with a Comptonization component (\textit{simpl}) in the energy range 0.7$-$30.0~keV. The spectral analysis revealed that the source was in its high soft state ($\Gamma$~=~2.67$^{+0.24}_{-0.24}$ and $\Gamma$~=~2.12$^{+0.19}_{-0.20}$) with a hot disc (kT$_{in}$~=~0.86$^{+0.01}_{-0.01}$ and kT$_{in}$~=~0.87$^{+0.02}_{-0.02}$). Thermal disc emission was fit with a relativistic model (\textit{kerrbb}) and spin of the black hole was estimated to be 0.93$^{+0.01}_{-0.01}$ and 0.93$^{+0.04}_{-0.03}$ (statistical errors) for the two \textit{Epochs} through X-ray continuum-fitting, which agrees with the previous results.
\end{abstract}
\section{Introduction}
LMC X-1, the first catalogued extra-galactic X-ray source in the Large Magellanic Cloud, was discovered in 1969 by \textit{Uhuru} \citep{Mar69,Pri71}. Since then, the source has been observed by most of the leading X-ray missions including the \textit{AstroSat} and \textit{NICER} and is persistently luminous (L$_{bol}$~=~2.2$\times$10$^{38}$~erg\ps) \citep{Lon81}. This binary system, located at a distance of 48.10$\pm$2.22~kpc \citep{Oro07,Oro09}, comprises of a black hole and an optical giant (O~III) companion star. While a canonical black hole binary undergoes spectral state transitions between hard and soft states, LMC X-1 has always been observed in the high soft state \citep{Wil01,Now01}. Transition to the hard state has never been seen.
\\[6pt]
\citet{Oro09} precisely determined the black hole mass (10.91$\pm$1.41~M$_{\odot}$), binary orbital inclination (36.38$\pm$1.92$^{\circ}$) and orbital period (3.90917$\pm$0.0005~d) using optical and near infrared data from the 1.3~m telescope at CTIO. Temporal analysis of \textit{XMM-Newton} data (0.3$-$10.0~keV) revealed the presence of a low frequency quasi periodic oscillation (QPO) at 26$-$29~mHz \citep{Ala14}. 
\\[6pt]
In the last two decades, there have been a few attempts to accurately determine spin of the black hole in LMC X-1. Indeed, such efforts are significant as the spin carries valuable information about the black hole formation mechanism and its growth history \citep{Mod98, Vol05}. X-ray reflection spectroscopy (analysis of the Fe-K~${\alpha}$ line) \citep{Bre06, Mil07,Rey14} and X-ray continuum-fitting (analysis of the disc continuum) \citep{Zhang97, Mcc14} are the two methods mostly used to estimate black hole spins as they are straightforward and involve very few parameters. \citet{Gie01} developed a general relativistic accretion disc model and applied it to constrain spin of the black hole in LMC X-1 by fitting X-ray continuum using soft X-ray data from \textit{ASCA} (0.7$-$10.0~keV). For the high-energy tail beyond the disc spectrum they chose a thermal Comptonization model \citep{Zdz96}. However, they could not give a good constraint on the black hole spin as the mass and disc inclination angle of LMC X-1 were evaluated rather poorly at that time. \citet{Gou09} determined the spin of the black hole in LMC X-1 to be 0.92$^{+0.05}_{-0.07}$ using eighteen thermal dominant spectra from \textit{RXTE$/$PCA} (3.0$-$30.0~keV) adopting X-ray continuum-fitting technique. They invoked the convolution model (\textit{simpl}) \citep{Ste09} to fit the Comptonization component, relativistic model (\textit{kerrbb}) \citep{Li05} to account for the thermal disc continuum emission and \textit{phabs} to correct for the interstellar absorption. \citet{Ste12} used 11 coordinated observations from \textit{RXTE$/$PCA-2} and \textit{Suzaku} (0.8$-$10.0~keV) to constrain the spin of the black hole in LMC X-1 adopting X-ray reflection spectroscopy. The spectrum was modelled using two different types of reflection models (\textit{refbhb} \citep{Ros05,Rei08}, and \textit{ireflect} \citep{Mag95} and a narrow \textit{Gaussian}) to account for the broad relativistic reflection features. Models \textit{kerrbb2} \citep{Li05,Dav06}, \textit{simpl} and \textit{tbvarabs} \citep{Wil00} were used in both the cases. The spin was found to be 0.97$^{+0.01}_{-0.13}$ and 0.97$^{+0.02}_{-0.25}$ for models involving \textit{refbhb} and \textit{ireflect}+\textit{Gaussian} respectively at 68 per cent confidence. \citet{Tri20} analysed seventeen observations from \textit{RXTE$/$PCA-2} for determining spin of the black hole in LMC X-1 employing the new model \textit{nkbb} developed by \citet{Zho19}. The spectra were modelled using a combination of \textit{nkbb} (thermal component), \textit{simpl} (Comptonization) and \textit{tbabs} (Galactic absorption). The black hole mass, distance and disc inclination angle were taken from \citet{Oro09} and spectral hardening factor was fixed to 1.55. The spin was constrained to be 0.998$_{-0.004}$ at 90 per cent confidence using X-ray continuum-fitting.
\\[6pt]
LMC X-1 is one of the few BHXBs whose mass, distance and binary orbital inclination angle are well constrained now. Many of the inferences of its black hole spin are primarily obtained from \textit{RXTE$/$PCA-2} data, which is limited by its low energy coverage (3.0$-$30.0~keV). However, as LMC X-1 has a characteristic temperature of $\sim$0.9~keV \citep{Ruh11} it is crucial to have a broadband coverage especially in the lower energy range ($<$3~keV) to obtain precise estimate of spin through spectral fitting. The Soft X-ray Telescope (SXT) \citep{Sin16,Sin17} and Large Area X-ray Proportional Counter (LAXPC) \citep{Yad16,Agr17} aboard the Indian Space Observatory \textit{AstroSat} \citep{Sin14} is ideally suited for this as it provides simultaneous broadband coverage (0.3$-$80.0~keV). In the recent work on the spectral evolution of GRS~1915+105 using simultaneous broadband (1.0$-$50.0~keV) data from \textit{AstroSat} by \citet{Mis20}, the importance of low energy coverage ($<$3~keV, SXT data) to constrain the inner disc radii and the use of convolution model \textit{simpl} to model the Comptonization component instead of commonly used power-law are well emphasised. Here, we report the detailed simultaneous broadband spectral analysis of the observations of LMC X-1 by SXT and LAXPC aboard \textit{AstroSat} and estimation of spin of the black hole.
\section{Data Reduction}
LMC X-1 was observed in its high soft state (HSS) by SXT and LAXPC on board \textit{AstroSat} on two occasions, separated by 9 months (Table~\ref{obsid}). Level 2 clean event files for all the orbits were obtained using AS1SXTLevel2-1.4b\footnote{\url{http://www.tifr.res.in/~astrosat$\_$sxt/sxtpipeline.html}} version of the official SXT pipeline from Level 1 photon counting mode data files of SXT. A single, exposure corrected, merged Level 2 clean event file was created using SXT event merger tool\footnote{\url{http://www.tifr.res.in/~astrosat$\_$sxt/dataanalysis.html}} by merging data from different orbits for each of the \textit{Epochs} separately and was used for the analysis.
\\[6pt]
The SXT spectrum was extracted using HEASoft version-6.24 tool XSELECT for a circular region of 16\arcmin~radius which includes $\sim$97 per cent of the total source photons. The response matrix file (RMF) ``sxt$\_$pc$\_$mat$\_$g0to12.rmf" and an off-axis auxiliary response file (ARF) generated using SXT ARF generation tool\footnote{\url{http://www.tifr.res.in/~astrosat$\_$sxt/dataanalysis.html}} appropriate for the source location on the charged coupled device (CCD) were used for the analysis. The blank sky observation(s): ``SkyBkg$\_$comb$\_$EL3p5$\_$Cl$\_$Rd16p0$\_$v01.pha" provided by the SXT POC team was used for background subtraction.
\\[6pt]
Level 2 event files were obtained from Level 1 event mode data of LAXPC using the latest version of LAXPC software\footnote{\url{http://astrosat-ssc.iucaa.in/?q=laxpcData}}. LAXPC subroutine 'laxpc$\_$make$\_$stdgti' \citep{Ant17} was used to generate good time interval (GTI) file to remove the Earth occultation of the source and South Atlantic Anomaly passes of the satellite. Total spectrum along with RMF for LAXPC10, LAXPC20 and LAXPC30 was obtained using ``laxpc$\_$make$\_$spectra". LAXPC background spectrum was extracted by employing the faint background model of LAXPC as the source has very low count rate in the energy range $>$30~keV. A lightcurve generated at 10~s time bin using data from all the three LAXPCs revealed that LAXPC30
was off for sometime during the observation. Therefore, data from LAXPC30 was not used in this work. Since LAXPC20 had lower background compared to LAXPC10, we used LAXPC20 data for the analysis.
\section{Spectral analysis}
Combined spectral fittings of SXT$-$LAXPC20 observations in the energy range 0.7$-$30.0~keV were carried out using XSPEC version-12.10.0c \citep{Arn96} with a combination of suitable models. While the higher energy regime ($>$30.0~keV) was ignored due to high background, the lower energy regime ($<$0.7~keV) was also not considered due to the uncertainties in effective area and response of the CCD of SXT. A systematic error of 3 per cent to account for uncertainties in response calibration and background uncertainty of 3 per cent were incorporated at the time of spectral fitting. 
\subsection{Modelling for the disc temperature and inner disc radius}
\begin{figure}
\includegraphics[width=0.47\textwidth]{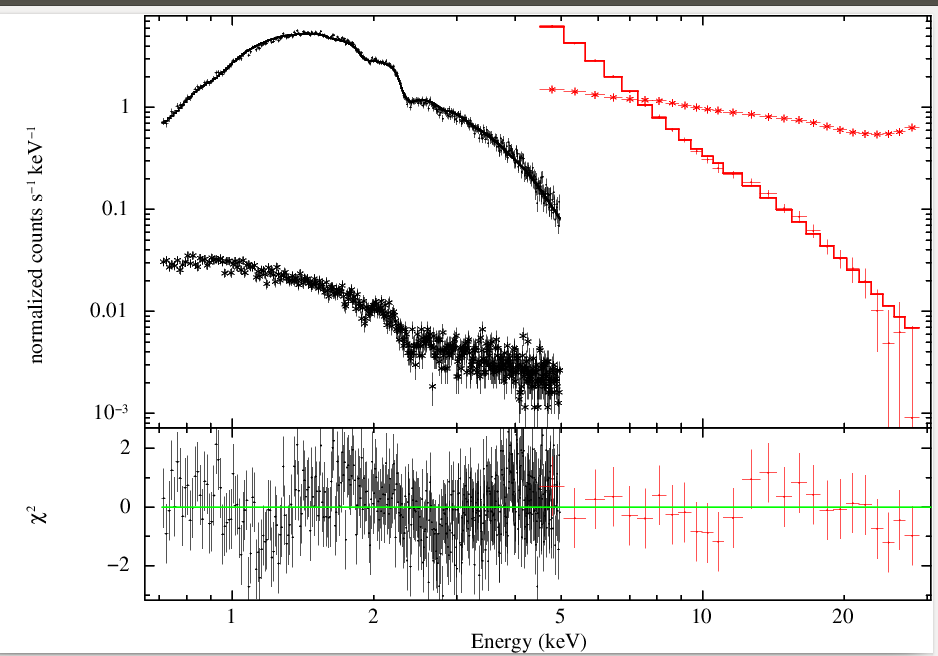}
\includegraphics[width=0.47\textwidth]{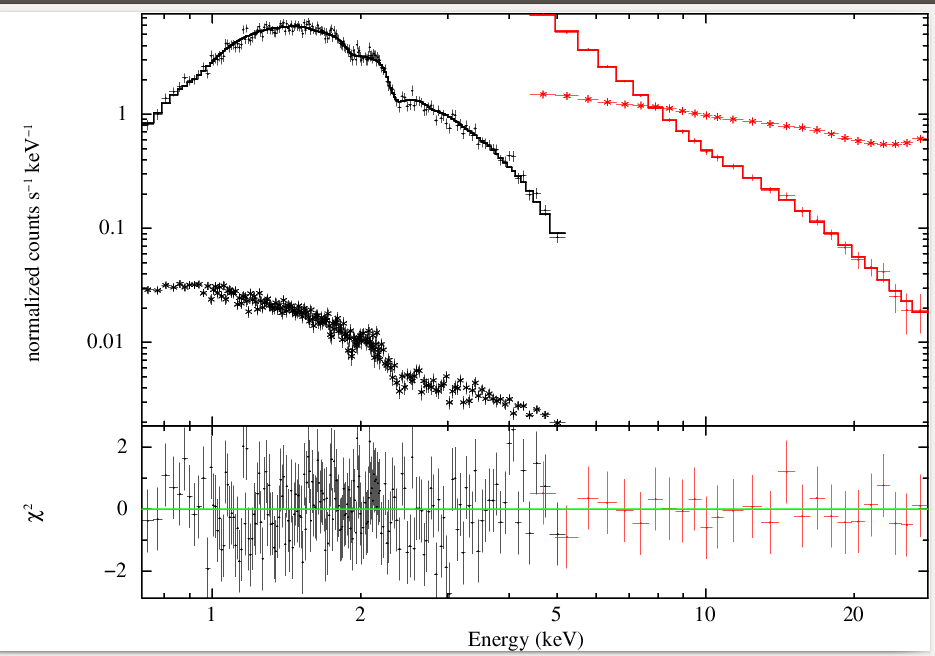}
\caption{SXT (0.7$-$5.0~keV) (black) and LAXPC20 (4.0$-$30.0~keV) (red) combined spectra with residuals for \textit{Epoch 1} (Top panel) and \textit{Epoch 2} (Bottom panel) fit with \textit{diskbb}, \textit{simpl}, \textit{Gaussian} and \textit{tbabs} models. The expected background spectra for SXT and LAXPC20 are also shown.}
\label{spec}
\end{figure}

\begin{figure}
\includegraphics[width=0.49\textwidth]{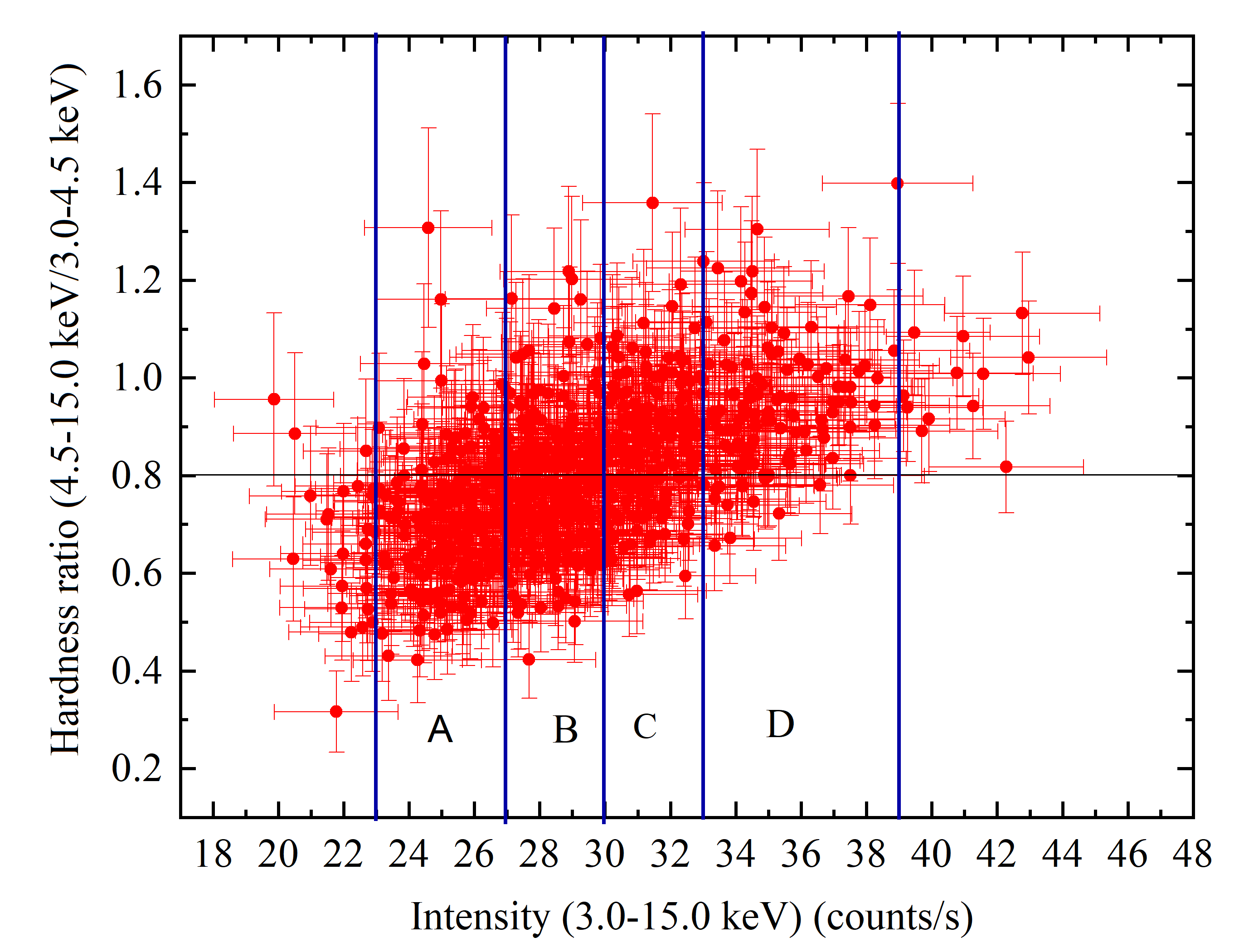}
\caption{HID of \textit{Epoch 1} binned at 10~s. Vertical lines indicate the regions A, B, C and D selected on the HID and horizontal line indicates the mean value of the hardness ratios.}
\label{hid}
\end{figure}

\begin{figure}
    \includegraphics[width=0.482\textwidth]{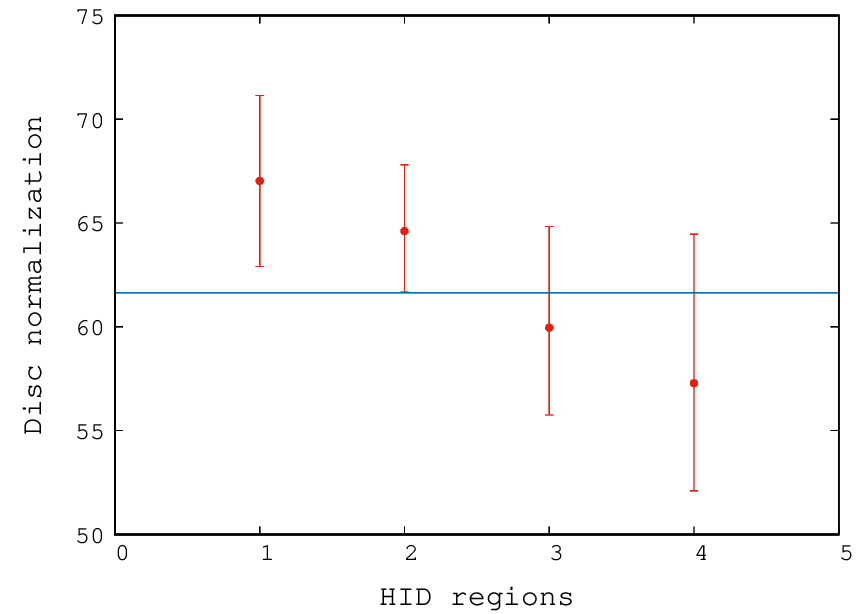}
    \caption{Normalization of diskbb at the regions A, B, C and D on HID of \textit{Epoch 1}. It should be noted that the number 1 on the X-axis corresponds to A, 2 corresponds to B and so on. The plot is fitted by a function f(x)~=~constant.}
    \label{disknorm}
\end{figure}

\begin{figure}
\includegraphics[width=0.50\textwidth]{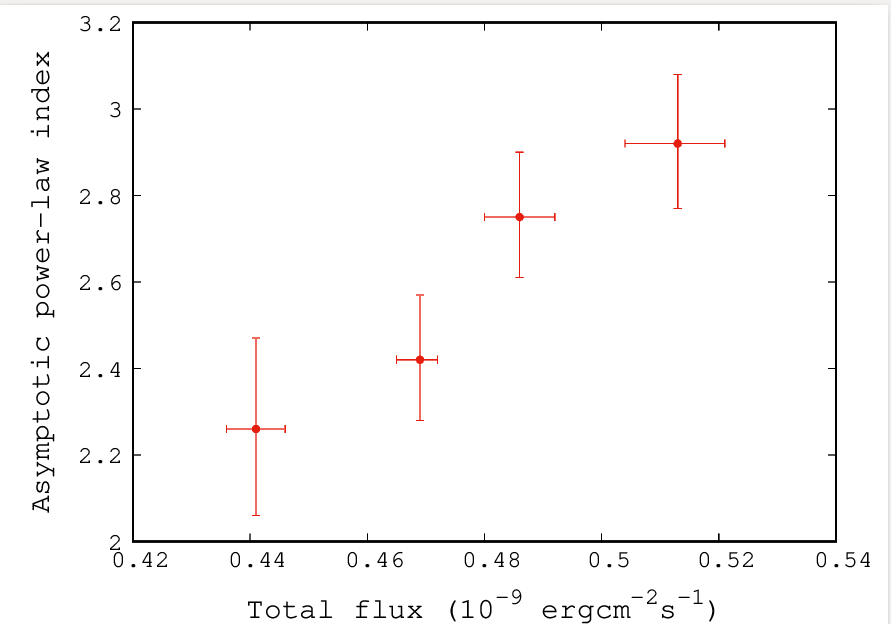}
\includegraphics[width=0.50\textwidth]{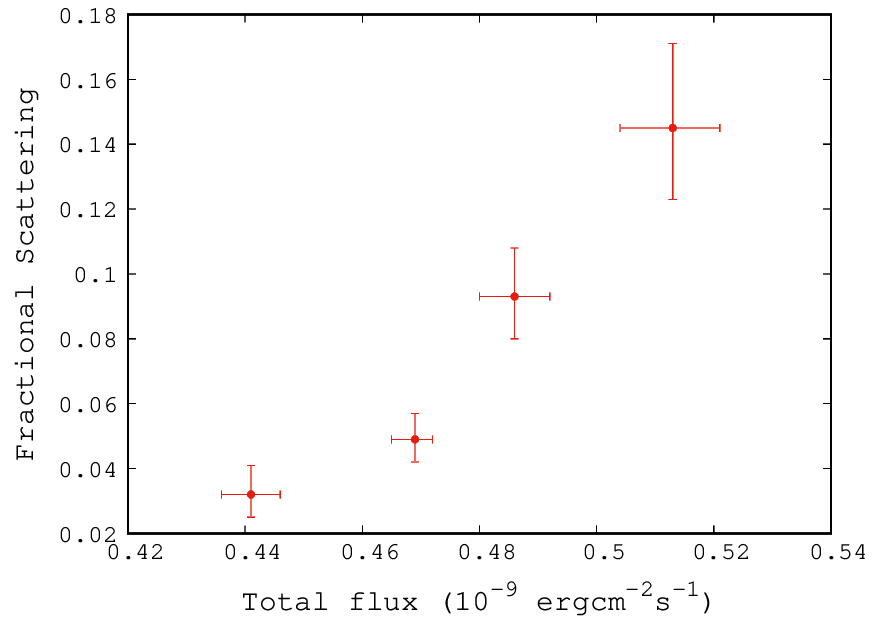}
\includegraphics[width=0.50\textwidth]{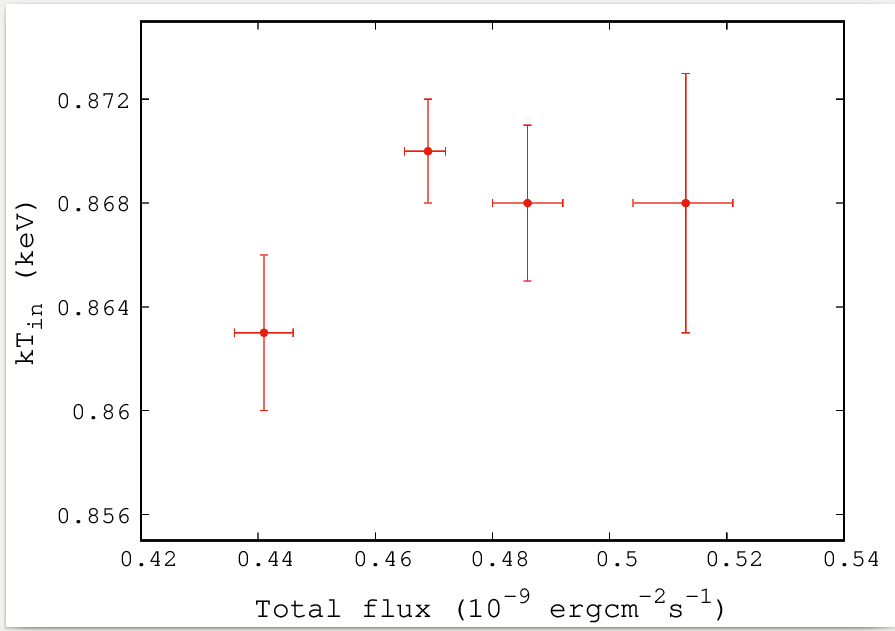}
\caption{The spectral fit parameters: Asymptotic power-law index (Top panel), Fractional scattering (Middle panel) and inner disc temperature (kT$_{in}$) (Bottom panel) as a function of total flux for the four regions on the HID with the disc normalization frozen at 61.64.}
\label{fluxres}
\end{figure}

The combined spectra having thermal and Comptonization components were fit with a combination of multicolour blackbody (\textit{diskbb}) \citep{Mit84,Mak86} and Comptonization (\textit{simpl}) models along with Galactic absorption (\textit{tbabs}). The relative normalization was allowed to vary between SXT and LAXPC20 spectra. The model showed some residuals at around 6.5~keV, suggesting the presence of an Iron line feature. Hence, we included a \textit{Gaussian} component with its centroid fixed at 6.5~keV allowing the width and normalization to vary. The additional component led to a $\Delta\chi^{2}$~$\sim$5 and $\sim$11 for \textit{Epoch 1} and \textit{2} respectively. Since SXT's effective area is relatively small at energies greater than 6~keV and LAXPC spectral resolution is limited with 3 per cent systematics, we refrained from using more complex relativistic models for the Iron line feature.
\\[6pt]
In \textit{Epoch 1}, the asymptotic power-law index, $\Gamma$~=~2.67$^{+0.24}_{-0.24}$, indicated that the source was in its HSS. Modelling the disc component revealed the disc temperature (kT$_{in}$) and normalization (\textit{N$_{disc}$}) to be 0.86$^{+0.01}_{-0.01}$~keV and 60.26$^{+4.84}_{-4.41}$ respectively. The best fit spectrum is shown in Figure~\ref{spec} (Top panel) and the best fit parameters are summarised in Table~\ref{spectab}. An offset gain correction of 32.6~eV determined using XSPEC command \textit{gainfit} with the slope fixed at unity was applied to the SXT data. The total flux and disc flux in the 0.7$-$30.0~keV energy band were estimated using the convolution model \textit{cflux} and the disc fraction was found to be $\sim$77 per cent of the total flux.
\begin{table}
\caption{Observations of LMC X-1}
\begin{tabular}{l c c c l l}
\toprule 
Obs. ID & Obs. Start Date & Obs. Start Time	& Exposure Time (ks)\\
 & yyyy-mm-dd & hh:mm:ss & SXT$~~~$LAXPC20 \\
\hline 
\\
9000000826 & 2016-11-26 & 13:39:42 &  29.86$~~~~~~~$44.5 \\
	   \\
9000001496  & 2017-08-28 & 19:56:23 & 3.414$~~~~~~~$15.7 \\
\bottomrule
\end{tabular}
\label{obsid}
\end{table}
\\[6pt]
Hardness intensity diagram (HID) was generated from the background subtracted LAXPC20 lightcurve at  10~s binning. We define the hardness as the ratio of count rate in the energy bands 4.5$-$15.0~keV and 3.0$-$4.5~keV and the intensity as the count rate in the energy band 3.0$-$15.0~keV. Four regions (A, B, C and D as shown in Figure~\ref{hid}) were selected from the HID such that each region had sufficient intensity. The hardness ratio was found to vary between 0.4$-$1.2 from region A to D. The GTI file appropriate for each region was retrieved using the LAXPC subroutine ``laxpc$\_$fluxresl$\_$gti''. The corresponding SXT and LAXPC spectra were then extracted using these GTI files. The resulting spectra were modelled using the same combination of models as before. A fixed offset gain correction of 32.6~eV was applied to the SXT data and the relative normalization between SXT and LAXPC was fixed at 1.17$^{+0.05}_{-0.05}$. The hydrogen column density was fixed at 0.46$\times$10$^{22}$\pcm. The Gaussian line energy, width and normalization were frozen at 6.5~keV, 1.0~keV and 2.33$\times$10$^{-4}$ respectively. The best fit parameters with errors are listed in Table~\ref{hidregtab}. As shown in Figure~\ref{disknorm}, the disc normalization at these four regions of the HID were almost a constant within  the uncertainties, indicating no substantial change in the inner disc radius of the accretion disc. Thus, we fitted these four nomalization values to a constant function and obtained a best fit value of 61.64. We subsequently fixed the disc normalization for all four spectra to this value and refitted the spectra. The resultant variation of the asymptotic power-law index, fractional scattering and inner disc temperature (kT$_{in}$) with total flux are plotted in Figure~\ref{fluxres}.
\\[6pt]
The \textit{Epoch 2} best fit spectrum is shown in Figure~\ref{spec} (Bottom panel) and the best fit parameters are given in Table~\ref{spectab}. An offset gain correction of 36.8~eV was applied to the SXT data. The disc temperature and normalization were found to be 0.87$^{+0.02}_{-0.02}$~keV and 66.76$^{+10.88}_{-9.30}$ respectively. The disc flux was found to be $\sim$84 per cent of the total flux. The HID for \textit{Epoch 2} did not show any variations. 
\begin{table*}
    \caption{The best fit parameters of SXT and LAXPC20 spectra - disc temperature and inner disc radius.}
    \begin{tabular}{cclcccccccc}
    \toprule
        \textit{Epoch} & constant & N$_{H}$  & Gamma ($\Gamma$)& FractSctr & Width & N$_{Gaussian}$ & kT$_{in}$ & N$_{disc}$ & $\chi{^2}/$dof \\
        & & (10$^{22}$~\pcm) & & & (keV) & (10$^{-4}$) & (keV)& &\\ 
       \\ 
         & Relative & Hydrogen & Asymptotic & Scattered frac- & Gaussian & Gaussian & Temperature & Disc &\\
         & norma-& column & power-law & tion of the  & width & norma- & at inner & norma- & \\
         & lization & density & index & seed photons & & lization & disc radius & lization &\\
         \midrule
         \\
        1 & 1.22$^{+0.06}_{-0.05}$ & 0.46$^{+0.01}_{-0.01}$ & 2.67$^{+0.24}_{-0.24}$ & 0.079$^{+0.029}_{-0.022}$ & $<$1.56 & 2.33$^{+2.31}_{-1.94}$ & 0.86$^{+0.01}_{-0.01}$ & 60.26$^{+4.84}_{-4.41}$ & 362$/$364 \\
         \\
         2 & 1.28$^{+0.08}_{-0.08}$ & 0.49$^{+0.02}_{-0.02}$ & 2.12$^{+0.19}_{-0.20}$ & 0.054$^{+0.016}_{-0.014}$ & 1.25$^{+0.42}_{-0.41}$ & 6.57$^{+3.09}_{-2.82}$ & 0.87$^{+0.02}_{-0.02}$ & 66.76$^{+10.88}_{-9.30}$ & 172$/$202 \\
         \\
         \toprule
    \end{tabular}
    \label{spectab}
\end{table*}
\begin{table*}
 \caption{Spectral fit parameters of four selected regions on the HID of the \textit{Epoch 1}}
 \begin{tabular}{l l l l l}
 \toprule
 &  & HID regions & & \\
 \hline
 & \textbf{A} & \textbf{B} & \textbf{C} & \textbf{D}\\
 \midrule
 \\
 Total flux (10$^{-9}$~erg\pcms) & 0.444$^{+0.005}_{-0.005}$ & 0.471$^{+0.004}_{-0.003}$ & 0.490$^{+0.005}_{-0.005}$ &  0.517$^{+0.007}_{-0.007}$\\
 \\
 SXT exposure time (s) & 2792 & 8906 & 2594 & 1020 \\
 \\
 LAXPC exposure time (s) & 8200 & 17000 & 8300 & 3400 \\
 \\
 Gamma ($\Gamma$) & 2.51$^{+0.30}_{-0.28}$ &  2.54$^{+0.19}_{-0.19}$ &  2.69$^{+0.21}_{-0.21}$ & 2.80$^{+0.24}_{-0.23}$ \\
 \\
 FractSctr & 0.045$^{+0.019}_{-0.013}$ & 0.057$^{+0.015}_{-0.011}$ & 0.087$^{+0.025}_{-0.019}$ & 0.125$^{+0.042}_{-0.031}$ \\
 \\
 kT$_{in}$ (keV) & 0.84$^{+0.14}_{-0.15}$ & 0.86$^{+0.01}_{-0.01}$ & 0.87$^{+0.02}_{-0.02}$ & 0.88$^{+0.02}_{-0.03}$ \\
 \\
 N$_{disc}$ & 67.03$^{+4.72}_{-4.12}$ & 64.61$^{+3.20}_{-2.92}$ & 59.96$^{+4.87}_{-4.21}$ & 57.29$^{+7.17}_{-5.78}$ \\
 \\
 $\chi{^2}/$dof & 167$/$167 & 273$/$280 & 151$/$165 & 82$/$84 \\
 \\
 \bottomrule
 \end{tabular}
    \label{hidregtab}
\end{table*}
\begin{table*}
    \caption{The best fit parameters of SXT and LAXPC20 spectra - spin of the black hole.}
    \begin{tabular}{ l l l l l l l l l l l l}
    \toprule
         \textit{Epoch} & constant & N$_{H}$ & Gamma ($\Gamma$)& FractSctr & Width & N$_{Gaussian}$ & a & $\dot{M}$  & L$^{(*)}$ & $\chi{^2}/$dof \\
       & & (10$^{22}$~\pcm)& & & (keV) &(10$^{-4}$) & & (10$^{18}$~g\ps) &\\
          & Relative & Hydrogen & Asymptotic& Scattered frac- & Gaussian & Gaussian & spin  & Rate of & Eddington\\
         & normali-  & column & power-law & tion of the & width & norma- & of the & accretion & ratio \\
         & zation & density & index & seed photons & &lization & black hole & &\\
         \midrule
         \\
         1 & 1.22$^{+0.04}_{-0.04}$ & 0.50 $^{+0.01}_{-0.01}$ & 2.40$^{+0.20}_{-0.19}$ & 0.049$^{+0.015}_{-0.011}$ & 1.0$^{(f)}$ & $<$0.87 & 0.93$^{+0.01}_{-0.01}$ & 1.17 $^{+0.08}_{-0.08}$ & 0.13$^{+0.01}_{-0.01}$ & 367$/$363 \\
         \\
        2 & 1.30$^{+0.06}_{-0.06}$ & 0.53 $^{+0.02}_{-0.02}$ & 2.03$^{+0.17}_{-0.10}$ & 0.044$^{+0.012}_{-0.011}$ & 1.35$^{+0.51}_{-0.58}$ & 3.98$^{+2.56}_{-2.29}$ & 0.93$^{+0.04}_{-0.03}$ & 1.39 $^{+0.20}_{-0.17}$ & 0.15$^{+0.02}_{-0.02}$ & 214$/$203 \\
         \\
         \bottomrule
    \end{tabular}
    \begin{flushleft}
    $^{(f)}$ Parameter frozen during the fitting. \\
    $^{(*)}$ L~=~L$_{acc}/$L$_{Edd}$, where L$_{acc}$~=~$\eta\dot Mc^{2}$ with $\eta\sim$0.17 for a~=~0.93 and L$_{Edd}$~=~1.3$\times$10$^{38}$(M$_{BH}/$M$_{\odot})$~erg\ps, M$_{BH}$~=~10.91$\pm$1.41~M$_{\odot}$
    \end{flushleft}
    \label{bhspectab}
\end{table*}
\subsection{Modelling for the black hole spin}
The combined spectrum of SXT$-$LAXPC20 (0.7$-$30.0~keV) was modelled using the combination of \textit{tbabs}, \textit{simpl}, \textit{Gaussian} and relativistic (\textit{kerrbb}) models to determine the spin of the black hole. The black hole mass (10.91~M$_{\odot}$) and source distance (48.10~kpc) were taken from \cite{Oro07, Oro09}. Furthermore, we assumed the disc inclination (i) to be same as the binary orbital inclination (36.38$^{\circ}$). The spectral hardening factor was frozen at 1.7 \citep{Shi95} and the \textit{kerrbb} normalization was fixed at 1. The Gaussian line energy was fixed at 6.5~keV for both the \textit{Epochs}. The Gaussian normalization for \textit{Epoch 1} was fixed at 1 whereas the width was allowed to vary. Meanwhile, for \textit{Epoch 2} both the Gaussian normalization and width were treated as variables.
\\[6pt]
SXT data required an offset gain of 43.4~eV (\textit{Epoch 1}) and 33.9~eV (\textit{Epoch 2}) to be added during the analysis. The best fit values of the spin and other parameters in both the \textit{Epochs} are presented in Table \ref{bhspectab}. The hydrogen column density and rate of accretion in \textit{Epoch 2} was found to be slightly higher than that in \textit{Epoch 1}, whereas the fractional scattering was found to be slightly lower in \textit{Epoch 2}. 
\begin{figure}
\includegraphics[width=0.48\textwidth]{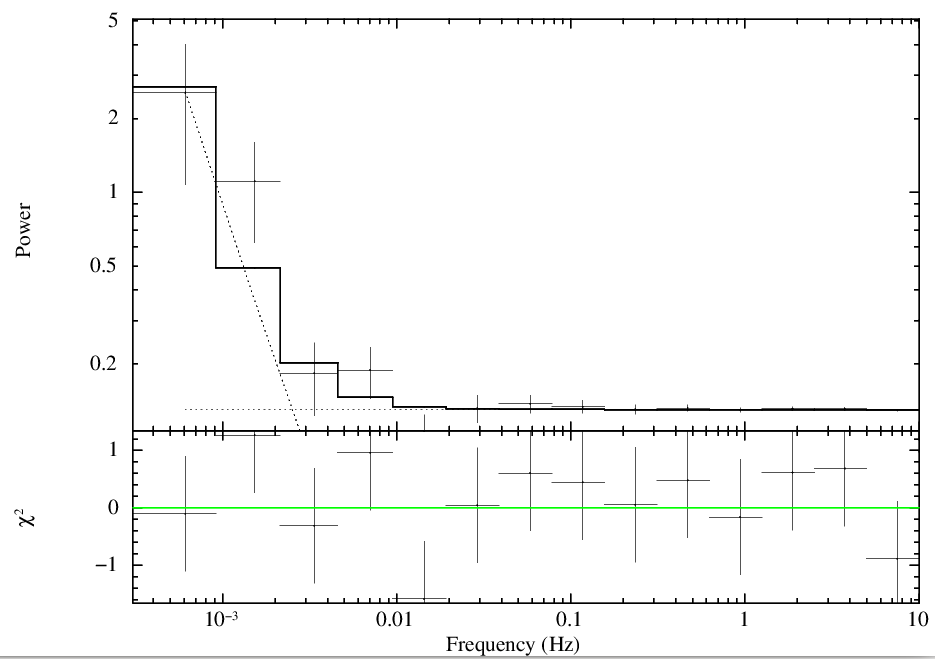}
\caption{Power density spectrum in the energy range 4.0$-$30.0~keV (Top panel) and residuals (Bottom panel) fitted by a power-law to model the low frequency excess and a constant to model the Poisson noise.} 
\label{pds}
\end{figure}
\begin{figure}
    \includegraphics[width=0.48\textwidth]{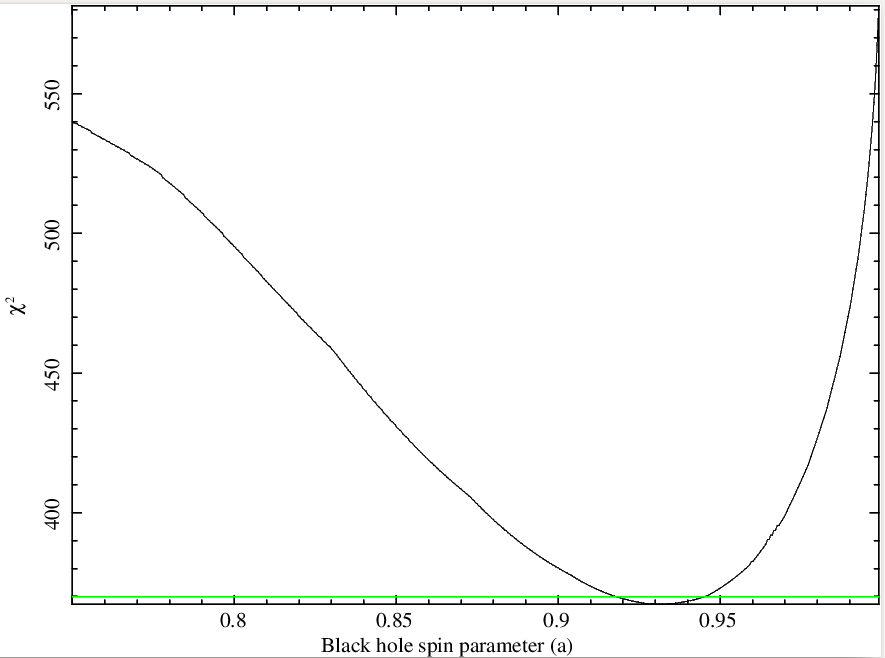}
    \caption{Variation of $\chi^{2}$ for different values of spin parameter (a), corresponding to the model parameters listed in Table~\ref{bhspectab}. Free parameters listed in Table~\ref{bhspectab} were allowed to vary while computing $\chi^{2}$. Spin parameter values of $<$0.8 are rejected with high significance.}
    \label{spin}
\end{figure}
\begin{figure}
     \includegraphics[width=0.48\textwidth]{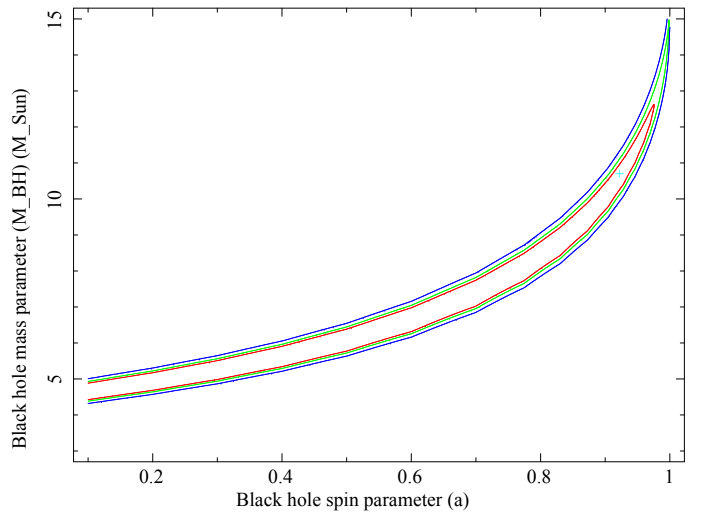}
     \includegraphics[width=0.48\textwidth]{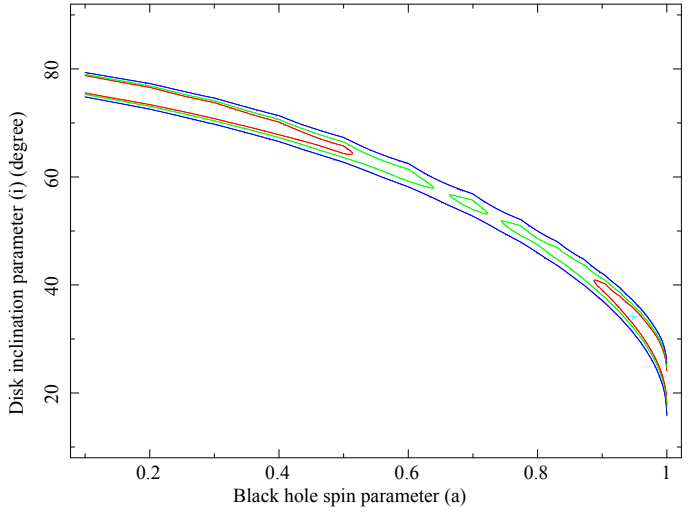}
     \includegraphics[width=0.48\textwidth]{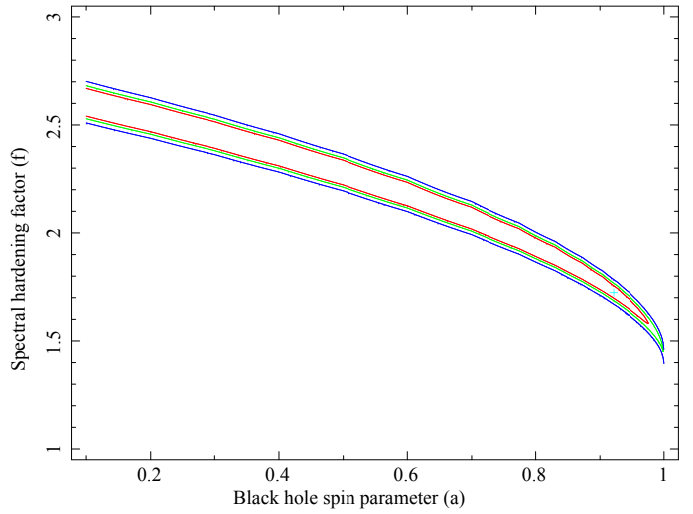}
      \caption{Contour plot of spin and mass of the black hole (Top panel), spin and disc inclination angle (i) (Middle panel) and spin and spectral hardening factor (f) (Bottom panel). The 68, 90 and 99 per cent contours are shown. The plus mark corresponds to the local minima (best fit) found by the fit.}
      \label{amif}
\end{figure}
\section{Timing analysis}
The complete event mode data of LAXPC in the energy range 4.0$-$30.0~keV was considered for the temporal analysis for the longer observation \textit{Epoch 1}. The lightcurve from LAXPC10 and LAXPC20 in the energy range 4.0$-$30.0~keV, with a time resolution of 0.05~s was divided into three segments of 1639.34~s each. The power spectra of the three segments were averaged to give a mean power density spectrum (PDS). The PDS was generated in the frequency range 0.6~mHz$-$10~Hz. Gaps in the data set prevented using larger time segments and hence the analysis was restricted to $>$0.6~mHz. The PDS  binned logarithmically in frequency is shown in Figure~\ref{pds}. At high frequencies the power is a constant due to Poisson noise, while some excess is detected at low frequencies, but no QPO like features are visible. Fitting the PDS with a power-law (to model the low frequency excess) and a constant (for the Poisson noise) yielded a $\chi^2/$dof of 9$/$12 and the best fit power-law index was found to be 2.01$^{+1.61}_{-0.68}$.
\section{Summary}
In this work, we have performed a comprehensive spectral analysis of LMC X-1, a persistent X-ray source hosting a Kerr black hole and an O III companion star, using the \textit{AstroSat} data, observed on 2016 November 26$^{th}$ (\textit{Epoch 1}) and 2017 August 28$^{th}$ (\textit{Epoch 2}) from SXT and LAXPC instruments.   
\\[6pt]
The combined SXT$-$LAXPC20 spectra were modelled with a combination of \textit{diskbb}, \textit{simpl}, \textit{Gaussian} and \textit{tbabs} models. Asymptotic power-law index for \textit{Epoch 1} ($\Gamma$~=~2.67$^{+0.24}_{-0.24}$) was slightly higher than \textit{Epoch 2} ($\Gamma$~=~2.12$^{+0.19}_{-0.20}$), whereas the disc temperature (0.86$^{+0.01}_{-0.01}$~keV) estimated at the inner disc radius of \textit{Epoch 1} was found to be slightly lower than that of \textit{Epoch 2} (0.87$^{+0.02}_{-0.02}$~keV). The hydrogen column density (N$_{H}$) increases slightly from \textit{Epoch 1} to \textit{Epoch 2} and N$_{H}$ determined at \textit{Epoch 1} agrees with the values of \citet{Ala14} and \citet{Cui02}.
\\[6pt]
Flux resolved spectroscopy performed for \textit{Epoch 1} revealed that the diskbb normalization remained almost a constant across the four regions (A, B, C, and D) on the HID. The fractional scattering increased from region A to D with a maximum value of $\sim$0.145 and a minimum of $\sim$0.032. The inner disc temperature varied from $\sim$0.863~keV (A) to $\sim$0.870~keV (D).
\\[6pt]
We invoked the relativistic model (\textit{kerrbb}) along with \textit{Gaussian} and \textit{tbabs} to fit the joint spectra of SXT and LAXPC20  and estimated the spin of the black hole (0.93$^{+0.01}_{-0.01}$ and 0.93$^{+0.04}_{-0.03}$ for \textit{Epoch 1} and \textit{Epoch 2} respectively). These values agree with the literature values \citep{Gou09, Ste12}. The uncertainties in the spin value reflect only the statistical errors during spectral fitting and do not include uncertainty in the distance and inclination measurements. If we allow distance and inclination to vary within their errors during spectral fitting, the spin parameter values have a larger allowed range i.e. 0.95$^{ +0.03}_ {-0.06}$ (\textit{Epoch 1}) and 0.93$^{ +0.05}_ {-0.06}$ (\textit{Epoch 2}). The relativistic accretion disc model, "\textit{kerrbb}" assumes that the disc is geometrically thin, which is valid when the luminosity is less than 0.3 times the Eddington value L$_{Edd}$ \citep{Mcc06}. As shown by \citet{Gou09}, the luminosity of LMC X-1 is indeed sub-Eddington. Here for \textit{Epoch 1} and \textit{Epoch 2}, we obtain the luminosity to be $\sim$0.13~L$_{Edd}$ and $\sim$0.15~L$_{Edd}$, thereby validating the use of "\textit{kerrbb}" to model the spectra.
\\[6pt]
Figure~\ref{spin} shows the variation of $\chi^{2}$ as a function of the black hole spin parameter (a). Low values of spin parameter (a~$<$0.8) are statistically rejected with high significance. We note again, that the spin parameter has been well determined because of the tight estimates on the mass of the black hole and binary orbital inclination angle. We illustrate this by showing the confidence levels for the spin parameter and black hole mass in the top panel of Figure~\ref{amif}. If the mass was unknown, at 2-$\sigma$ level, the spin parameter would have been only constrained to $>$0.2. However, as illustrated in the confidence level plots for inclination angle and spin parameter in the middle panel of Figure~\ref{amif}, if the inclination angle was not known, the spin parameter would not have been constrained at all. In this work, we have assumed the spectral hardening factor to be the expected theoretical value, f~=~1.7 \citep{Shi95}. We investigated the influence of varying this parameter on the spin measurement. The bottom panel of Figure~\ref{amif} shows the confidence contours for f parameter and black hole spin by keeping the black hole mass and inclination angle fixed. We note that the data restricts f to be within 1.4$-$2.1 at 1-$\sigma$ level, which includes the expected value of 1.7. Moreover, the spin parameter is constrained to be greater than 0.7, if the spectral hardening factor is less than 2.0.
\\[6pt]
The LAXPC10 and LAXPC20 lightcurve in energy range 4.0$-$30.0~keV, with a time resolution of 0.05~s was generated from the complete event mode data of LAXPC. The lightcurve was divided into three segments of 1639.34~s each. The power spectra of the three segments were averaged to give a mean PDS. The PDS was generated in the frequency range 0.6~mHz$-$10~Hz. We noticed that the power remained constant due to Poisson noise at higher frequencies and showed some excess at lower frequencies. However, no QPO like features were seen.
\\[6pt]
In summary, we have estimated the spin of the persistent black hole binary LMC X-1 using SXT and LAXPC data of \textit{AstroSat} through X-ray continuum-fitting. Future observations of the source by \textit{AstroSat} and simultaneous observations by \textit{NICER} in the soft band and \textit{NuSTAR} in the hard band, can be used to confirm the results presented here and provide perhaps more stringent constraints on the spin parameters. With spin and other non varying parameters of the system well constrained, spectral modelling will provide a good measure of the accretion rate and parameters of the thermal plasma producing the Comptonized component. Frequent monitoring of the source would provide a unique opportunity to study how these parameters evolve with time and the relation between them.
\clearpage
\section*{Acknowledgements}
The anonymous referee is thanked for the valuable comments and suggestions which greatly helped in improving this manuscript. This work has used the data from the Soft X-ray Telescope (SXT) developed at TIFR, Mumbai, and the SXT POC at TIFR is thanked for verifying and releasing the data via the ISSDC data archive and providing the necessary software tools. The Large Area X-ray Proportional Counter POC team is also thanked for their support and timely release of data. This work has made use of software provided by HEASARC. SPM thanks CHRIST (Deemed to be University), Bengaluru for the Research Assistant position through MRPDSC-1721. The authors (SBG $\&$ BSG) acknowledge Centre for Research, CHRIST (Deemed to be University), Bengaluru for the research funding (MRPDSC-1721). SBG thanks the Inter-University Centre for Astronomy and Astrophysics (IUCAA), Pune for Visiting Associateship. AR acknowledges a Commonwealth Rutherford Fellowship.
\section*{Data availabilty}
The data utilised in this article are available at \textit{AstroSat}$-$ISSDC website (\url{http://astrobrowse.issdc.gov.in/astro$\_$archive/archive/Home.jsp)}. The software used for data analysis is available at NASA's HEASARC website (\url{https://heasarc.gsfc.nasa.gov/lheasoft/download.html)}.
\newpage
\bibliographystyle{mnras}
\bibliography{bibliography.bib}

\begin{thebibliography}{}
\makeatletter
\relax
\def\mn@urlcharsother{\let\do\@makeother \do\$\do\&\do\#\do\^\do\_\do\%\do\~}
\def\mn@doi{\begingroup\mn@urlcharsother \@ifnextchar [ {\mn@doi@}
  {\mn@doi@[]}}
\def\mn@doi@[#1]#2{\def\@tempa{#1}\ifx\@tempa\@empty \href
  {http://dx.doi.org/#2} {doi:#2}\else \href {http://dx.doi.org/#2} {#1}\fi
  \endgroup}
\def\mn@eprint#1#2{\mn@eprint@#1:#2::\@nil}
\def\mn@eprint@arXiv#1{\href {http://arxiv.org/abs/#1} {{\tt arXiv:#1}}}
\def\mn@eprint@dblp#1{\href {http://dblp.uni-trier.de/rec/bibtex/#1.xml}
  {dblp:#1}}
\def\mn@eprint@#1:#2:#3:#4\@nil{\def\@tempa {#1}\def\@tempb {#2}\def\@tempc
  {#3}\ifx \@tempc \@empty \let \@tempc \@tempb \let \@tempb \@tempa \fi \ifx
  \@tempb \@empty \def\@tempb {arXiv}\fi \@ifundefined
  {mn@eprint@\@tempb}{\@tempb:\@tempc}{\expandafter \expandafter \csname
  mn@eprint@\@tempb\endcsname \expandafter{\@tempc}}}

\bibitem[\protect\citeauthoryear{{Agrawal}}{{Agrawal}}{2017}]{Agr17}
{Agrawal} P.~C.,  2017, JApA, 38, 27

\bibitem[\protect\citeauthoryear{{Alam}, {Dewangan}, {Belloni}, {Mukherjee}  \&
  {Jhingan}}{{Alam} et~al.}{2014}]{Ala14}
{Alam} M.~S.,  {Dewangan} G.~C.,  {Belloni} T.,  {Mukherjee} D.,   {Jhingan}
  S.,  2014, \mnras, 445, 4259

\bibitem[\protect\citeauthoryear{{Antia} et~al.,}{{Antia} et~al.}{2017}]{Ant17}
{Antia} H.~M.,  et~al., 2017, \apjs, 231

\bibitem[\protect\citeauthoryear{{Arnaud}}{{Arnaud}}{1996}]{Arn96}
{Arnaud} K.~A.,  1996, {XSPEC: The First Ten Years}.
Astronomical Data Analysis Software and Systems V, eds. Jacoby G. and Barnes
  J., p.~17

\bibitem[\protect\citeauthoryear{{Brenneman} \& {Reynolds}}{{Brenneman} \&
  {Reynolds}}{2006}]{Bre06}
{Brenneman} L.~W.,  {Reynolds} C.~S.,  2006, \apj, 652, 1028

\bibitem[\protect\citeauthoryear{{Cui}, {Feng}, {Zhang}, {Bautz}, {Garmire}  \&
  {Schulz}}{{Cui} et~al.}{2002}]{Cui02}
{Cui} W.,  {Feng} Y.~X.,  {Zhang} S.~N.,  {Bautz} M.~W.,  {Garmire} G.~P.,
  {Schulz} N.~S.,  2002, \apj, 576, 357

\bibitem[\protect\citeauthoryear{{Davis} \& {Hubeny}}{{Davis} \&
  {Hubeny}}{2006}]{Dav06}
{Davis} S.~W.,  {Hubeny} I.,  2006, \apjs, 164, 530

\bibitem[\protect\citeauthoryear{{Gierli{\'n}ski},
  {Macio{\l}ek-Nied{\'z}wiecki}  \& {Ebisawa}}{{Gierli{\'n}ski}
  et~al.}{2001}]{Gie01}
{Gierli{\'n}ski} M.,  {Macio{\l}ek-Nied{\'z}wiecki} A.,   {Ebisawa} K.,  2001,
  \mnras, 325, 1253

\bibitem[\protect\citeauthoryear{{Gou} et~al.,}{{Gou} et~al.}{2009}]{Gou09}
{Gou} L.,  et~al., 2009, \apj, 701, 1076

\bibitem[\protect\citeauthoryear{{Li}, {Zimmerman}, {Narayan}  \&
  {McClintock}}{{Li} et~al.}{2005}]{Li05}
{Li} L.-X.,  {Zimmerman} E.~R.,  {Narayan} R.,   {McClintock} J.~E.,  2005,
  \apjs, 157, 335

\bibitem[\protect\citeauthoryear{{Long}, {Helfand}  \& {Grabelsky}}{{Long}
  et~al.}{1981}]{Lon81}
{Long} K.~S.,  {Helfand} D.~J.,   {Grabelsky} D.~A.,  1981, \apj, 248, 925

\bibitem[\protect\citeauthoryear{{Magdziarz} \& {Zdziarski}}{{Magdziarz} \&
  {Zdziarski}}{1995}]{Mag95}
{Magdziarz} P.,  {Zdziarski} A.~A.,  1995, \mnras, 273, 837

\bibitem[\protect\citeauthoryear{{Makishima}, {Maejima}, {Mitsuda}, {Bradt},
  {Remillard}, {Tuohy}, {Hoshi}  \& {Nakagawa}}{{Makishima}
  et~al.}{1986}]{Mak86}
{Makishima} K.,  {Maejima} Y.,  {Mitsuda} K.,  {Bradt} H.~V.,  {Remillard}
  R.~A.,  {Tuohy} I.~R.,  {Hoshi} R.,   {Nakagawa} M.,  1986, \apj, 308, 635

\bibitem[\protect\citeauthoryear{{Mark}, {Price}, {Rodrigues}, {Seward}  \&
  {Swift}}{{Mark} et~al.}{1969}]{Mar69}
{Mark} H.,  {Price} R.,  {Rodrigues} R.,  {Seward} F.~D.,   {Swift} C.~D.,
  1969, \apjl, 155, L143

\bibitem[\protect\citeauthoryear{{McClintock}, {Shafee}, {Narayan},
  {Remillard}, {Davis}  \& {Li}}{{McClintock} et~al.}{2006}]{Mcc06}
{McClintock} J.~E.,  {Shafee} R.,  {Narayan} R.,  {Remillard} R.~A.,  {Davis}
  S.~W.,   {Li} L.-X.,  2006, \apj, 652, 518

\bibitem[\protect\citeauthoryear{{McClintock}, {Narayan}  \&
  {Steiner}}{{McClintock} et~al.}{2014}]{Mcc14}
{McClintock} J.~E.,  {Narayan} R.,   {Steiner} J.~F.,  2014, \ssr, 183, 295

\bibitem[\protect\citeauthoryear{{Miller}}{{Miller}}{2007}]{Mil07}
{Miller} J.~M.,  2007, \araa, 45, 441

\bibitem[\protect\citeauthoryear{Misra, Rawat, Yadav  \& Jain}{Misra
  et~al.}{2020}]{Mis20}
Misra R.,  Rawat D.,  Yadav J.~S.,   Jain P.,  2020, \apjl, 889, L36

\bibitem[\protect\citeauthoryear{{Mitsuda} et~al.,}{{Mitsuda}
  et~al.}{1984}]{Mit84}
{Mitsuda} K.,  et~al., 1984, \pasj, 36, 741

\bibitem[\protect\citeauthoryear{{Moderski}, {Sikora}  \& {Lasota}}{{Moderski}
  et~al.}{1998}]{Mod98}
{Moderski} R.,  {Sikora} M.,   {Lasota} J.~P.,  1998, \mnras, 301, 142

\bibitem[\protect\citeauthoryear{{Nowak}, {Wilms}, {Heindl}, {Pottschmidt},
  {Dove}  \& {Begelman}}{{Nowak} et~al.}{2001}]{Now01}
{Nowak} M.~A.,  {Wilms} J.,  {Heindl} W.~A.,  {Pottschmidt} K.,  {Dove} J.~B.,
   {Begelman} M.~C.,  2001, \mnras, 320, 316

\bibitem[\protect\citeauthoryear{{Orosz} et~al.,}{{Orosz} et~al.}{2007}]{Oro07}
{Orosz} J.~A.,  et~al., 2007, \nat, 449, 872

\bibitem[\protect\citeauthoryear{{Orosz} et~al.,}{{Orosz} et~al.}{2009}]{Oro09}
{Orosz} J.~A.,  et~al., 2009, \apj, 697, 573

\bibitem[\protect\citeauthoryear{{Price}, {Groves}, {Rodrigues}, {Seward},
  {Swift}  \& {Toor}}{{Price} et~al.}{1971}]{Pri71}
{Price} R.~E.,  {Groves} D.~J.,  {Rodrigues} R.~M.,  {Seward} F.~D.,  {Swift}
  C.~D.,   {Toor} A.,  1971, \apjl, 168, L7

\bibitem[\protect\citeauthoryear{{Reis}, {Fabian}, {Ross}, {Miniutti}, {Miller}
   \& {Reynolds}}{{Reis} et~al.}{2008}]{Rei08}
{Reis} R.~C.,  {Fabian} A.~C.,  {Ross} R.~R.,  {Miniutti} G.,  {Miller} J.~M.,
   {Reynolds} C.,  2008, \mnras, 387, 1489

\bibitem[\protect\citeauthoryear{{Reynolds}}{{Reynolds}}{2014}]{Rey14}
{Reynolds} C.~S.,  2014, \ssr, 183, 277

\bibitem[\protect\citeauthoryear{{Ross} \& {Fabian}}{{Ross} \&
  {Fabian}}{2005}]{Ros05}
{Ross} R.~R.,  {Fabian} A.~C.,  2005, \mnras, 358, 211

\bibitem[\protect\citeauthoryear{{Ruhlen}, {Smith}  \& {Swank}}{{Ruhlen}
  et~al.}{2011}]{Ruh11}
{Ruhlen} L.,  {Smith} D.~M.,   {Swank} J.~H.,  2011, \apj, 742, 75

\bibitem[\protect\citeauthoryear{{Shimura} \& {Takahara}}{{Shimura} \&
  {Takahara}}{1995}]{Shi95}
{Shimura} T.,  {Takahara} F.,  1995, \apj, 445, 780

\bibitem[\protect\citeauthoryear{{Singh} et~al.,}{{Singh} et~al.}{2014}]{Sin14}
{Singh} K.~P.,  et~al., 2014, in \procspie. p. 91441S

\bibitem[\protect\citeauthoryear{{Singh} et~al.,}{{Singh} et~al.}{2016}]{Sin16}
{Singh} K.~P.,  et~al., 2016, in \procspie. p. 99051E

\bibitem[\protect\citeauthoryear{{Singh} et~al.,}{{Singh} et~al.}{2017}]{Sin17}
{Singh} K.~P.,  et~al., 2017, JApA, 38, 29

\bibitem[\protect\citeauthoryear{{Steiner}, {Narayan}, {McClintock}  \&
  {Ebisawa}}{{Steiner} et~al.}{2009}]{Ste09}
{Steiner} J.~F.,  {Narayan} R.,  {McClintock} J.~E.,   {Ebisawa} K.,  2009,
  \pasp, 121, 1279

\bibitem[\protect\citeauthoryear{{Steiner} et~al.,}{{Steiner}
  et~al.}{2012}]{Ste12}
{Steiner} J.~F.,  et~al., 2012, \mnras, 427, 2552

\bibitem[\protect\citeauthoryear{{Tripathi} et~al.,}{{Tripathi}
  et~al.}{2020}]{Tri20}
{Tripathi} A.,  et~al., 2020, arXiv e-prints, p. arXiv:2001.08391

\bibitem[\protect\citeauthoryear{{Volonteri}, {Madau}, {Quataert}  \&
  {Rees}}{{Volonteri} et~al.}{2005}]{Vol05}
{Volonteri} M.,  {Madau} P.,  {Quataert} E.,   {Rees} M.~J.,  2005, \apj, 620,
  69

\bibitem[\protect\citeauthoryear{{Wilms}, {Allen}  \& {McCray}}{{Wilms}
  et~al.}{2000}]{Wil00}
{Wilms} J.,  {Allen} A.,   {McCray} R.,  2000, \apj, 542, 914

\bibitem[\protect\citeauthoryear{{Wilms}, {Nowak}, {Pottschmidt}, {Heindl},
  {Dove}  \& {Begelman}}{{Wilms} et~al.}{2001}]{Wil01}
{Wilms} J.,  {Nowak} M.~A.,  {Pottschmidt} K.,  {Heindl} W.~A.,  {Dove} J.~B.,
   {Begelman} M.~C.,  2001, \mnras, 320, 327

\bibitem[\protect\citeauthoryear{{Yadav} et~al.,}{{Yadav} et~al.}{2016}]{Yad16}
{Yadav} J.~S.,  et~al., 2016, in \procspie. p. 99051D

\bibitem[\protect\citeauthoryear{{Zdziarski}, {Johnson}  \&
  {Magdziarz}}{{Zdziarski} et~al.}{1996}]{Zdz96}
{Zdziarski} A.~A.,  {Johnson} W.~N.,   {Magdziarz} P.,  1996, \mnras, 283, 193

\bibitem[\protect\citeauthoryear{Zhang, Cui  \& Chen}{Zhang
  et~al.}{1997}]{Zhang97}
Zhang S.~N.,  Cui W.,   Chen W.,  1997, The Astrophysical Journal, 482, L155

\bibitem[\protect\citeauthoryear{{Zhou}, {Abdikamalov}, {Ayzenberg}, {Bambi},
  {Liu}  \& {Nampalliwar}}{{Zhou} et~al.}{2019}]{Zho19}
{Zhou} M.,  {Abdikamalov} A.~B.,  {Ayzenberg} D.,  {Bambi} C.,  {Liu} H.,
  {Nampalliwar} S.,  2019, \prd, 99, 104031

\makeatother
\end{thebibliography}
\bsp	
\label{lastpage}
\end{document}